\documentclass[a4paper,10pt,twocolumn,superscriptaddress,nofootinbib]{revtex4}
\usepackage{amsmath,amssymb}
\usepackage[final]{graphicx}
\usepackage[english]{babel}
\usepackage[utf8]{inputenc}
\usepackage{hyperref}

\graphicspath{{./}}
\makeatletter
\renewcommand{\p@subsection}{}
\renewcommand{\p@subsubsection}{}
\makeatother

    \def\<{\langle}
    \def\>{\rangle}

    \def\vphi{\varphi}
    \def\pt{\partial}

  \newcommand{\eq}[1]{
    \begin{equation}
    {#1}
    \end{equation}}

    \newcommand{\nmq}[1]{
    \begin{multline}
    #1
    \end{multline}}

\begin{document}
\title{Scaling dimension of quantum Hall quasiparticles from tunneling current noise measurements}
\author{Kyrylo Snizhko}
\affiliation{Physics Department, Taras Shevchenko National University of Kyiv, Kyiv, 03022, Ukraine}
\affiliation{Department of Physics, Lancaster University, Lancaster, LA1~4YB, UK}
\author{Vadim Cheianov}
\affiliation{Department of Physics, Lancaster University, Lancaster, LA1~4YB, UK}

\begin{abstract}
Determination of properties of quasiparticle excitations is an important task in the experimental investigation of the fractional quantum Hall effect (FQHE). We propose a model-independent method for finding the scaling dimension of FQHE quasiparticles from measurements of the electric current tunneling between two FQHE edges and its noise. In comparison to the commonly used method based on measuring the tunneling current only, the proposed method is less prone to the errors due to non-universal physics of tunnel junctions.
\end{abstract}

\maketitle

\section*{INTRODUCTION}

The fractional quantum Hall effect (FQHE, fractional QHE) has long been known to occur due to electrons forming a strongly correlated topologically ordered state \cite{Wen_TopOrder}. While the bulk of this state has a gap, gapless excitations are always present at the FQHE edge. The excitation spectrum and dynamical properties of these edge modes can be encoded in an effective low-energy theory. Such theories, called Chiral Luttinger Liquids (CLL) \cite{WenReview}, provide a powerful theoretical framework for the description of the fractional quantum Hall effect (FQHE). However, for a given filling factor $\nu$ there may exist several candidate theories predicting the same value of the Hall conductance, but possessing different excitation spectra (e.g., they may differ by whether non-Abelian quasiparticles are present or by the number of transport channels). Probably, the most studied example of such a variety of models is the much debated $\nu = 5/2$ state \cite{HalperinRosenow_2007_AntiPfaffian, Nayak_2007_AntiPfaffian, BoCheFro_RaduExp_analysis}, first observed in the distant 1987 \cite{Willett_EvenDenomQHObserv}. In such a situation, the foremost task in the investigation of the FQHE state is to discriminate between the candidate theories.

An important characteristic of an edge theory is the spectrum of local quasiparticle excitations. Each quasiparticle is characterized by several quantum numbers, of which two are important for the present paper: the electric charge and the scaling dimension. These quantum numbers can, in principle, be determined in experiments involving tunneling of quasiparticles between two FQHE edges. In this article we discuss weak quasiparticle tunneling through the FQHE bulk in a quantum point contact (QPC). In this case the quasiparticle with the smallest scaling dimension (the most relevant quasiparticle) gives the most important contribution to transport. One can hope to extract the charge and scaling dimension of the particle from transport measurements in such a system. Even such a limited amount of data as the properties of the most relevant quasiparticle can significantly reduce the number of candidate theories. This can be seen, for example, from the theoretical study of Ref.~\cite{BoCheFro_RaduExp_analysis}, relating to the $\nu = 5/2$ state.

It is, in principle, possible to extract the charge and the scaling dimension from the tunneling current measurements only (see the experimental work of Refs.~\cite{Radu_experiment, Ensslin_FQHE_TunnExp} and references to theory therein). Though, it is well known (see e.g. \cite{TunnellingRate0, TunnellingRate1, TunnellingRate2} and references therein) that the dependence of the tunneling current on the applied bias voltage in electrostatically confined QPCs strongly disagrees with the predictions of theoretical models. Even in the simplest FQHE case of $\nu = 1/3$ experimental and theoretical curves agree only qualitatively but not quantitatively (see e.g. Ref.~\cite{Exp_Lauglin_TunnConduct}).\footnote{Moreover, in the case of $\nu = 1$ the experimental curves also deviate from the behaviour one would expect theoretically \cite{IQHE_TunRate_Experiment}. Ref.~\cite{IQHE_TunRate_Experiment} explains this by emergence of isles of fractional QHE in the QPC region. However, phenomenologically one can interpret this as the bias voltage dependence of the tunneling barrier form that determines tunneling between the $\nu = 1$ edges.} One of the simplest possible explanations of this is that tunneling amplitudes depend on the applied bias voltage in an unknown non-universal way due to electrostatic effects resulting in changing of the tunneling barrier form. Other possible explanations tell that the tunneling operator scaling dimension gets renormalized due to (a) interaction with additional degrees of freedom (such as phonon modes \cite{Rosenow2002} or $1/f$ noise \cite{Carrega2011}), (b) Coulomb interaction between different parts of the FQHE edge at the QPC \cite{Papa2004}. For the most of this paper we stick to the former explanation, we briefly discuss the effects of other explanations on our results in section~\ref{Sect_Discussion}.

Thus, the charge and scaling dimension extracted from the measurements of tunneling current only are prone to large systematic errors. Another popular observable to look at is the tunneling current noise and related quantities (see e.g. \cite{Bena2007, Ferraro2014, HeiblumExp, Dolev2008}). However, analysing noise only is also prone to errors due to non-universal physics at the QPC.

The natural solution is to analyse the tunneling current and its noise simultaneously which would allow one to exclude some of the non-universal physics. Indeed, it is analysing the noise and the tunneling current together (namely their ratio at large bias voltages~--- the Fano factor) that allowed confirmation of the fractional charge of quasiparticles in the $\nu = 1/3$ FQHE for the first time \cite{FracChargeObs_Heiblum, FracChargeObs_Glattli}. Only several years later was the fractional charge in the $\nu = 1/3$ state observed by another method \cite{Martin2004}. Also for more complicated states like $\nu = 2/3$, considering the ratio of the tunneling current noise to the tunnleing current allows one to reduce significantly the influence of non-universal physics at the QPC (assuming that the non-universal physics results in an unknown dependence of the tunneling amplitudes on the applied bias voltage, but not in the scaling dimension renormalization) \cite{ShtaSniChe_2014}.

In this paper we focus on the possibility to extract the scaling dimension of the most relevant quasiparticle from such a ratio, paying particular attention to the $\nu = 1/3$ case as the simplest one. The structure of the paper is as follows. In section~\ref{Sect_ScalDimFromAsymp} we introduce the model we are considering and show how the data on the tunneling current noise and the tunneling can be used to extract the tunneling quasiparticle scaling dimension. Then in section~\ref{Sect_ExactResults} we compare our results, which are perturbative in quasiparticle tunneling, with the exact solution available for $\nu = 1/3$. Finally, in section~\ref{Sect_Discussion} we discuss the experimental conditions necessary to extract the scaling dimension using our method (subsection~\ref{SubSect_ExpConditions}) and how our method is going to be affected by the effects recently predicted or reported (subsection~\ref{SubSect_Relations}).

\section{\label{Sect_ScalDimFromAsymp}Scaling dimension from the noise to tunneling rate ratio}

\begin{figure}[tbp]
  \center{\includegraphics[width=1\linewidth]{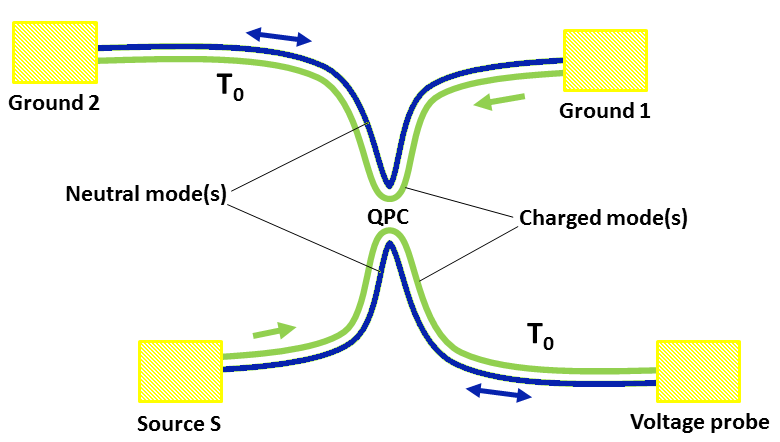}}
  \caption{\textbf{A realistic setup of tunneling current noise measuring experiment}. (Color online). Current $I_s$ injected from the Ohmic contact \textit{Source S} to the lower edge charged mode(s), changing its (their) chemical potential(s). Tunneling of the quasiparticles at the constriction induces extra noise in the charged mode of the lower edge which is detected at the \textit{Voltage probe}.}
  \label{fig_exp_setup}
\end{figure}

We consider the following experimental setup (see Fig.~\ref{fig_exp_setup}). There are two quantum Hall edges, each supporting the same set of excitation modes. By an excitation mode we mean a channel in which long-lived excitations propagate in one direction with the same velocity. We call the propagation direction of a mode its chirality. The set of excitation modes includes a charge carrying mode or, possibly, several such modes, all having the same chirality, and some (possibly, none) neutral modes (that is the modes that do not carry electric charge). Neutral modes can have different chiralities. The two edges are far apart from each other except for the quantum point contact region where they come close to each other and quasiparticle tunneling processes take place. Yellow rectangles are the Ohmic contacts, which absorb everything that flows into them. Contacts \textit{Ground 1} and \textit{Ground 2} are grounded. Contact \textit{Source S} is used to inject direct electric current (DC) $I_s$ into the lower edge, and contact \textit{Voltage probe} is used to measure the electric current flowing into it and the current noise at zero frequency. All components of the system have absolute temperature~$T_0$.

In Ref.~\cite{ShtaSniChe_2014} we discussed a framework that allows one to deal with such experiments. Now we briefly recall it, and then we write out our results. For simplicity, we concentrate on the case of the Abelian FQHE edge models. However, our results can be readily generalized for a wide class of non-Abelian FQHE edge models, including the ones corresponding to the Moore-Read Pfaffian and anti-Pfaffian (in the disorder-dominated phase \cite{HalperinRosenow_2007_AntiPfaffian, Nayak_2007_AntiPfaffian}) states.\footnote{See Appendix~\ref{App_HowGeneralNoiseAnswers} for a detailed discussion of this issue.}

A single edge (the upper one for definiteness) is described by $N$ bosonic fields $\vphi_i$ with $i = 1,...,N$, each representing an edge mode. In the absence of electric field along the edge the fields can be described by the action
\eq{\label{action}
S = \frac1{4\pi} \int dx dt \sum_m \Bigl( -\chi_m \partial_x\vphi_m \partial_t \vphi_m - v_{m}(\partial_x\vphi_m)^2 \Bigl),
}
where $\chi_m = \pm1$ represent chiralities of the modes (plus for left-movers and minus for right-movers direction), and $v_{m} > 0$ are the modes' propagation velocities. Without loss of generality, we put $\chi_m = +1$ for $m = 1,..., N_{l}$ and $\chi_m = -1$ for $m = N_{l}+1, ..., N$ ($N_l$ is thus the number of left-moving modes).

The electric current $J^\mu$ ($J^0$ is the electric charge density, $J^1$ is the electric current flowing along the edge) has the form
\eq{\label{el_current operator_2} J^\mu = \frac{1}{2\pi}\sum_m q_m\varepsilon^{\mu\nu}\pt_\nu\vphi_m,}
where the symbol $\varepsilon^{\mu\nu}$ denotes the fully antisymmetric tensor with $\mu, \nu$ taking values $t$ and $x$ (or 0 and 1 respectively) and $\varepsilon^{t x} = \varepsilon^{0 1} = 1$. The numbers $q_i$ should satisfy the constraint \cite{WenCLL, FrohlichAnomaly}
\eq{\label{anomaly_cancellation}\sum_m \chi_m q_m^2 = \nu.}

We restrict our consideration to the case $q_m = 0$ for $m = N_{l}+1, ..., N$, i.e. only left-propagating modes can carry electric charge.\footnote{Our reasons for such a restriction are briefly discussed in Appendix~\ref{App_HowGeneralNoiseAnswers}.}

The quantized fields $\vphi_m$ obey the commutation relations
\eq{\label{comm_relations}[\vphi_m(x,t),\vphi_{m'}(x',t')] = -i \pi \mathrm{sgn}(X_m-X_m')\,\delta_{m,m'},}
where $X_m = -\chi_m x+v_m t$.

The edge supports quasiparticles of the form
\eq{\label{vertex_quasi}V_{\mathbf{g}}(x,t) = \left( \frac{L}{2\pi} \right)^{-\sum_m g_m^2/2} :\exp\left(i\sum_m g_m\vphi_m(x,t)\right):,}
which are important for the processes of tunneling at the \textit{QPC} (L is the edge length). The notation $: ... :$ stands for the normal ordering, $\mathbf{g} = (g_1,...,g_N)$, and $g_m\in\mathbb{R}$ are the quasiparticle quantum numbers. The quasiparticles' quantum numbers are quantized, i.e. the set of allowed vectors $\mathbf{g}$ is discrete. The quasiparticle's two most important quantum numbers, the electric charge $Q$ and the scaling dimension $\delta$, are equal to
\begin{eqnarray}
\label{Charge}
Q &=& \sum_m \chi_m q_m g_m,\\
\label{ScalDim}
\delta &=& \frac 12\sum_m g_m^2.
\end{eqnarray}

We model the QPC by the following Hamiltonian that describes tunneling of quasiparticles between the two edges:
\eq{\label{tun_ham} H_{T} = \sum_{\mathbf{g}} \eta_{\mathbf{g}} V_{\mathbf{g}}^{(u)\dag}(0,t)V^{(l)}_{\mathbf{g}}(0,t) + \mathrm{h.c.},}
where the superscripts $(u), (l)$ label quantities relating to the upper and the lower edge respectively, $\eta_{\mathbf{g}}$ are the tunneling amplitudes. Since both general theoretical arguments and Monte Carlo simulations \cite{HuntingtonCheianov_TunnAmplitudeMonteCarlo} suggest that the contributions of the quasiparticles with greater scaling dimensions are strongly suppressed at low enough energies, in our model the sum runs over the quasiparticles with the smallest scaling dimension $\delta$ only. In the following we label such quasiparticle types by $i = 1,...,n$, with the quasiparticle electric charges being $Q_i$ (in the units of the elementary charge $e$), their common scaling dimension being $\delta_i = \delta$, and the full set of quantum numbers being $\mathbf{g}^i$.

The current $I$ flowing into \textit{Voltage probe} contact (see Fig.~\ref{fig_exp_setup}) is equal to $J^1$ component of the current $J^\mu$, defined in Eq.~\eqref{el_current operator_2}, taken at some point to the right of the QPC along the lower edge. Let us denote the operator of this current as $\hat{I}(t)$. Then the average current flowing into \textit{Voltage probe} is $I = \langle \hat{I}(t) \rangle$. It is also convenient to introduce operator $\widehat{\delta I}(t) = \hat{I}(t) - I$.

If there is no tunneling at the QPC, the $I$ is equal to the current $I_s$ injected at \textit{Source S}. As soon as there is tunneling, some part of the quasiparticles will not reach \textit{Voltage probe} with $I_s - I = I_T$ being the tunneling current. We define two quantities:
\begin{itemize}
\item tunneling rate $r = I_T/I_s$;
\item measured current noise \eq{S(\omega) = \int\limits_{-\infty}^\infty d\tau \exp\bigl(i\omega\tau \bigl) \frac 12 \Bigl\<\Bigl\{\Delta I(0),\Delta I(\tau)\Bigl\}\Bigl\>,}
\end{itemize}
where $\{\dots\}$ denotes the anti-commutator, and \mbox{$\Delta I = I - \<I\>$}.

In what follows we only use the zero-frequency noise $S(\omega = 0)$. It is also convenient to talk about the excess noise
\eq{\tilde{S}(\omega = 0) = S(0) - S_{\mathrm{Nyquist}}(0) = S(0) - \frac{\nu}{2\pi} T_{0},}
where $T_0$ is the system temperature.

Applying the framework outlined above to the experimental setup described in the beginning of this section, in the lowest non-trivial order perturbation theory in the tunneling Hamiltonian \eqref{tun_ham}, we find the tunneling rate $r$ and the excess noise at zero frequency $\tilde{S}(\omega = 0)$\footnote{Here we restore the elementary charge $e$, the Planck constant $\hbar$, and the Boltzmann constant $k_B$, which we had put to $1$ in the formulae above.}:
\eq{\label{tunn_coeff_expr}r = \frac{4 e (\pi k_B T_0)^{4\delta - 1}}{I_s \hbar^{4 \delta + 1}} \sum_i \kappa_i G_i,}
\eq{\label{noise_expr}\tilde{S}(0) = \frac{4 e^2 (\pi k_B T_0)^{4\delta - 1}}{\hbar^{4 \delta + 1}} \sum_i \kappa_i F_i,}
\eq{
\label{G_i}
G_i = \sin{2\pi\delta} \int\limits_{0}^{\infty}dt \frac {Q_i \,\sin Q_i j_s t}{(\sinh t)^{4\delta}}
}
\eq{
\label{F_i}
F_i = F^{TT}_{i} \cos{2\pi\delta} - \frac{2}{\pi} F^{0T}_i \sin{2\pi\delta},
}
\eq{
\label{F_TT}
F^{TT}_{i} = Q_i^2 \lim_{\varepsilon \rightarrow +0} \left(\frac{\varepsilon^{1-4\delta}}{1-4\delta} + \int\limits_{\varepsilon}^{\infty}dt \frac{\cos Q_i j_s t}{(\sinh t)^{4\delta}} \right)
}
\eq{
\label{F_0T}
F^{0T}_{i} = \int\limits_{0}^{\infty}dt \frac {Q_i^2\, t\cos Q_i j_s t}{(\sinh t)^{4\delta}}
}
\eq{
\label{j_s}
j_s = \frac{I_s}{I_0},\quad I_0 = \nu \frac{e}{h} \pi k_B T_0,
}
where $T_0$ is the system temperature, $e$ is the elementary charge, $h = 2 \pi \hbar$ is the Planck constant, $k_B$ is the Boltzmann constant, $\nu$ is the filling factor, $\kappa_i = |\eta_{\mathbf{g}^i}|^2 \prod_m v_m^{-2 (g^i_m)^2}$, and $i$ enumerates different quasiparticles participating in tunneling. We remind the reader that $Q_i$ are the electric charges of the quasiparticles and $\delta$ is their common scaling dimension. The formulae \eqref{G_i}, \eqref{F_TT}, \eqref{F_0T} are correct for $\delta < 1/2$, for $\delta \geq 1/2$ they should be modified. However, typically the quasiparticles contributing to the tunneling processes are predicted to have $\delta < 1/2$.

In practice, experimental data rarely agrees with the predictions of Eq.~\eqref{tunn_coeff_expr} (see Refs.~\cite{TunnellingRate0, TunnellingRate1, TunnellingRate2, Exp_Lauglin_TunnConduct, IQHE_TunRate_Experiment}).\footnote{Although the data of Refs.~\cite{Radu_experiment, Ensslin_FQHE_TunnExp} agrees with similar theoretical predictions strikingly well, the values of quasiparticle charge $Q$ and scaling dimension $\delta$ do not coincide with the ones predicted by theory.} One of the simplest explanations of this is that the tunneling amplitudes, and therefore the parameters $\kappa_i$, have some unknown non-universal dependence on the current $I_s$, which complicates a comparison of experimental data with the theory. However, consideration of the ratio of the excess noise to the tunneling rate (NtTRR)
\eq{
\label{noise_tun}X(I_s) = \frac{\tilde{S}(0)}{r} = e I_s \frac{\sum_{i} \kappa_i F_i}{\sum_{i} \kappa_i G_i}
}
allows one to exclude the unwanted non-universal dependence in the case of one quasiparticle type dominating tunneling and reduce its influence in the case when several quasiparticles participate in the tunneling processes.\footnote{In the case of only one quasiparticle participating in tunneling the sum in both the numerator and the denominator consists of a single term and, therefore, the parameter $\kappa_1$ cancels out. A similar cancellation happens when the tunneling quasiparticles form a multiplet with the same electric charge and scaling dimension (so all the $F_i$ functions coincide, as do the $G_i$ functions). This happens, for example in the Jain states for $\nu = p/(2p+1)$, $p \geq 2$. In the case of several essentially different quasiparticles tunneling such a nice cancellation does not happen. However, by multiplying both the numerator and the denominator with $\kappa_1^{-1}$ one can show that the independent non-universal parameters that enter the NtTRR are not $\kappa_i$ but rather the ratios of $\kappa_i/\kappa_1$. Therefore, the number of independent non-universal parameters entering the NtTRR is reduced by one  compared to pure excess noise or pure tunneling current. Moreover, in our recent study \cite{ShtaSniChe_2014} we found that the experimental data of Ref.~\cite{HeiblumExp} regarding $\nu = 2/3$ can be described very well assuming that the ratios of $\kappa_i/\kappa_j$ are constant.}

Consider the large-$I_s$ limit of Eq.~\eqref{noise_tun}. For $|I_s| \gg I_0$ one gets\footnote{See Appendix~\ref{App_LargeCurrAsympDeriv} for derivation.}
\nmq{
\label{Noise_Tun_asymp_full}
X(I_s)\left|_{|j_s| \gg 1}\right. = e I_s \frac{\sum_{i} \kappa_i F_i}{\sum_{i} \kappa_i G_i} =\\
 = e |I_s| \frac{\sum_i \kappa_i Q_i^{4\delta+1}}{\sum_i \kappa_i Q_i^{4\delta}} + e I_0 \frac{2-8\delta}{\pi} + O\left(|j_s|^{-1}\right).
}
This is the key result of the present paper.

The leading term of the asymptotic behaviour \eqref{Noise_Tun_asymp_full} gives the well-known result that in the regime of weak tunneling the gradient of the noise to tunneling rate ratio is equal to the tunneling quasiparticle's charge. Here it is some average of the charges in the case of several quasiparticles participating in tunneling. Note the subleading term: constant offset contains information about the quasiparticles' scaling dimension. It is important that all the quasiparticles which significantly contribute to tunneling have the same scaling dimension.

Thus, in principle, by fitting large-$I_s$ experimental data with a linear function one can find not only the "effective charge" of the tunneling quasiparticles but also their scaling dimension (which is the same for all of the most relevant quasiparticles).

However, as we mentioned above, the parameters $\kappa_i$ related to the quasiparticles' tunneling amplitudes often depend on the current $I_s$ in a non-universal way. Therefore, Eq.~\eqref{Noise_Tun_asymp_full} is not useful in the case of several different quasiparticle charges as the gradient of the leading term may depend on $I_s$. From now on we concentrate on the case when all the charges of the quasiparticles contributing to tunneling are equal: $Q_i = Q$. Examples include the states of Laughlin series, the Moore-Read Pfaffian, Jain's $\nu = 2/5$ state etc. Then, independently of $\kappa_i$,
\eq{
\label{Noise_Tun_asymp_full_single_charge}
X(I_s)\left|_{|j_s| \gg 1}\right. = e Q |I_s| + e I_0 \frac{2-8\delta}{\pi} + O\left(|j_s|^{-1}\right).
}
Let us note that in this case it is possible to write a simple analytic expression for the NtTRR \eqref{noise_tun} (not just the large-$I_s$ asymptote)\footnote{Derivation is given in Appendix~\ref{App_AnalyticAnswerForNtTRR_Deriv}.}:
\eq{
\label{Noise_Tun_analyt_single_charge}
X(I_s)\left|_{Q_i = Q}\right. = \frac{2 e Q I_s}{\pi} \mathrm{Im}{\left[\psi{\left(2\delta + \frac{i Q j_s}{2}\right)}\right]},
}
where the digamma function $\psi{(x)} = (\ln{\Gamma{(x)}})'$ is the logarithmic derivative of the Euler gamma function $\Gamma{(x)}$, and $\mathrm{Im}{\left[...\right]}$ denotes taking of the imaginary part. Alongside the asymptotic expression \eqref{Noise_Tun_asymp_full_single_charge}, the full expression \eqref{Noise_Tun_analyt_single_charge} can also be used to extract the scaling dimension $\delta$ from experimental data.

We note that unlike Eqs.~\eqref{G_i}, \eqref{F_TT}, \eqref{F_0T}, results \eqref{Noise_Tun_asymp_full}, \eqref{Noise_Tun_asymp_full_single_charge}, \eqref{Noise_Tun_analyt_single_charge} are valid for any $\delta > 0$.

\section{\label{Sect_ExactResults}Exact results for $\nu = 1/3$ and the conditions to extract the scaling dimension by perturbative formulae}

In this section we concentrate on the filling factor $\nu = 1/3$, the simplest FQHE state. We compare our perturbative results presented in the previous section with the exact solution \cite{Saleur, Saleur2, Saleur_LaughlinNoise_detailed} available for this case in order to find the conditions when our results are reliable to use.

The minimal edge model for this filling factor has only one edge mode represented by the chiral bosonic field and can be constructed in the way described in section IV of Ref.~\cite{ShtaSniChe_2014}. The electric charge and the scaling dimension of the only most relevant quasiparticle in this model are respectively equal to $Q = 1/3$, $\delta = 1/6$.

This model is believed to give the correct description of the FQHE at $\nu = 1/3$. However, there is surprisingly little experimental evidence directly confirming this belief. While the charge of the most relevant quasiparticle has been confirmed long time ago \cite{FracChargeObs_Heiblum, FracChargeObs_Glattli, Martin2004}, this is not true for its statistics or other properties of the model.\footnote{Moreover, recently there has been a report \cite{Heiblum_2013_LaughlinFilling_NeutrMode}, results of which may be interpreted as a signature of presence of additional neutral modes in the $\nu = 1/3$ FQHE. However, in the present work we are not going to discuss this evidence.} Therefore, finding the most relevant quasiparticle's scaling dimension, using the method described in the previous section, would be an important check of validity of the minimal model.

We concentrate on the case of zero temperature of the system ($T_0 = 0$), for which analytic expressions are available. The finite temperature case requires solution of thermodynamic Bethe anzatz equations and is beyond the scope of this work.

The exact answer for the tunneling rate $r = I_T/I_s$ at zero temperature is as follows \cite{Saleur2}:
\eq{
\label{tun_rate_exact_Laughlin_1}
r\left(|I_s| > \Xi e^{\zeta}\right) = \nu \sum_{n = 1}^{\infty} A_n(\nu) \left(\frac{|I_s|}{\Xi}\right)^{2 n (\nu - 1)},
}
\nmq{
\label{tun_rate_exact_Laughlin_2}
r\left(|I_s| < \Xi e^{\zeta}\right) =\\ 1 - \nu^{-1} \sum_{n = 1}^{\infty} A_n(\nu^{-1}) \left(\frac{|I_s|}{\Xi}\right)^{2 n (\nu^{-1} - 1)},
}
\begin{eqnarray}
A_n(x) &=& (-1)^{n + 1} \frac{\sqrt{\pi} \Gamma{(n x)}}{2 \Gamma{(n)} \Gamma{(3/2 + n (x - 1))}},\\
\zeta &=& \frac{1}{2} \ln{\left(1 - \nu\right)} + \frac{\nu}{2 (1 - \nu)} \ln{\nu}.
\end{eqnarray}
The tunneling amplitude of the only tunneling quasiparticle $\eta$, the parameter $\kappa \propto |\eta|^2$ in the perturbative formulae \eqref{tunn_coeff_expr}, \eqref{noise_expr}, and the parameter $\Xi$ here are related: $\Xi \propto |\eta|^{1/(1-\nu)}$. Thus, $\Xi$ characterizes the tunneling strength. The restrictions on $|I_s|$ in the formulae \eqref{tun_rate_exact_Laughlin_1}, \eqref{tun_rate_exact_Laughlin_2} represent the radii of convergence of the series.\footnote{In Ref.~\cite{Saleur2} the definition of $\zeta$ (which is called $\Delta$ there) contains a misprint. However, one can check and find that the radius of convergence of the series leads to the definition of $\zeta$ presented here.}

According to Ref.~\cite{Saleur2}, at zero temperature the excess noise at zero frequency $\tilde{S}(\omega = 0)$ is connected to the tunneling rate $r$ via
\eq{
\tilde{S}(\omega = 0, I_s) = \frac{\nu e}{2(1-\nu)} |I_s| \times \Xi \frac{\partial}{\partial \Xi} r(I_s).
}
The explicit series are
\nmq{
\label{noise_exact_Laughlin_1}
\tilde{S}\left(\omega = 0, |I_s| > \Xi e^{\zeta}\right) =\\ \nu^2 e |I_s| \sum_{n = 1}^{\infty} n A_n(\nu) \left(\frac{|I_s|}{\Xi}\right)^{2 n (\nu - 1)},
}
\nmq{
\label{noise_exact_Laughlin_2}
\tilde{S}\left(\omega = 0, |I_s| < \Xi e^{\zeta}\right) =\\ \nu^{-1} e |I_s| \sum_{n = 1}^{\infty} n A_n(\nu^{-1}) \left(\frac{|I_s|}{\Xi}\right)^{2 n (\nu^{-1} - 1)}.
}

It is easy to recognize expansion in the orders of the tunneling amplitude $\eta$ in the formulae \eqref{tun_rate_exact_Laughlin_1}, \eqref{noise_exact_Laughlin_1}. Taking only the first term in the sums in Eqs.~\eqref{tun_rate_exact_Laughlin_1}, \eqref{noise_exact_Laughlin_1} one should recover the lowest order perturbation theory result for the regime of weak tunneling. This is indeed the case.\footnote{There is a small subtlety here. To adapt the perturbative answers \eqref{tunn_coeff_expr}-\eqref{j_s} for $T_0 = 0$ one should take the limit $T_0 \rightarrow 0$ which coincides with the limit $|j_s| >> 1$. Then up to a factor one recovers the expression one can get from taking only the first term in the sums in Eqs.~\eqref{tun_rate_exact_Laughlin_1}, \eqref{noise_exact_Laughlin_1}. This factor is related to the proportionality factor between $\Xi$ and $|\eta|^{1/(1-\nu)}$.} Note that while the perturbative NtTRR $X_{\mathrm{pert}}(I_s) = \tilde{S}_{\mathrm{pert}}\left(\omega = 0, I_s\right)/r_{\mathrm{pert}}\left(I_s\right)$ does not depend on the value of the tunneling amplitude $\eta$ (or $\Xi$, which is equivalent), the exact NtTRR does.

\begin{figure}[tbp]
  \center{\includegraphics[width=1\linewidth]{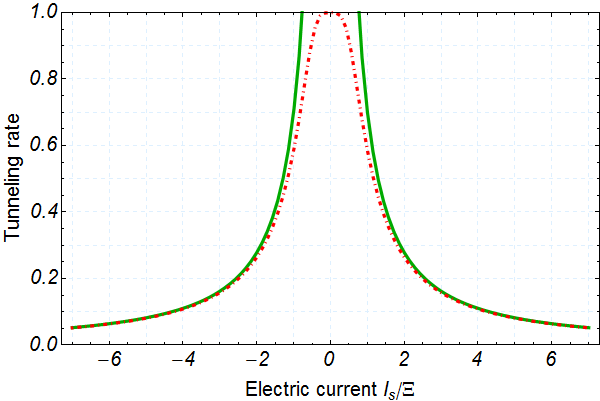}}
  \caption{\textbf{Tunneling rate at $\nu = 1/3$. Perturbative answer vs exact answer.} (Color online). The red dot-dashed curve is the exact tunneling rate given by Eqs.~\eqref{tun_rate_exact_Laughlin_1}, \eqref{tun_rate_exact_Laughlin_2}. The green solid curve is the lowest order perturbation theory answer for the tunneling rate, which can be obtained by taking only the first term in the sum in Eq.~\eqref{tun_rate_exact_Laughlin_1}. We remind the reader that the system temperature is equal to $T_0 = 0$.}
  \label{fig:Tun_rate_pert_VS_exact}
\end{figure}

We now compare the exact answers with the perturbative ones. Fig.~\ref{fig:Tun_rate_pert_VS_exact} shows the comparison of the perturbative and the exact answers for the tunneling rate. For tunneling rates\footnote{We remind that the tunneling rate lies between $0$ and $1$ by definition.} not exceeding $0.2$ the two answers are reasonably close. Note, that knowing the tunneling rate at a certain value of the current $I_s$ one can find the corresponding value of the tunneling amplitude $\Xi$.

\begin{figure}[tbp]
  \center{\includegraphics[width=1\linewidth]{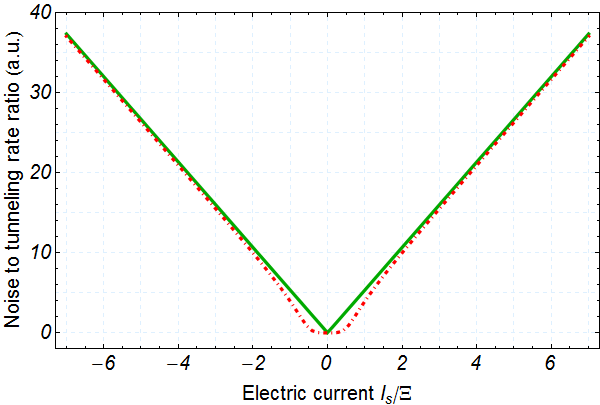}}
  \caption{\textbf{Noise to tunneling rate ratio at $\nu = 1/3$. Perturbative answer vs exact answer.} (Color online). The red dot-dashed curve is the exact NtTRR plotted using the Eqs.~\eqref{tun_rate_exact_Laughlin_1}--\eqref{noise_exact_Laughlin_2}. The green solid curve is the lowest order perturbation theory answer for the NtTRR, which can be obtained by taking only the first term in the sums in Eqs.~\eqref{tun_rate_exact_Laughlin_1}, \eqref{noise_exact_Laughlin_1}. We remind the reader that the system temperature is equal to $T_0 = 0$.}
  \label{fig:NtTRR_pert_VS_exact}
\end{figure}

Fig.~\ref{fig:NtTRR_pert_VS_exact} shows the comparison of the perturbative and the exact answers for the noise to tunneling rate ratio. Since the temperature $T_0 = 0$ the perturbative answer for NtTRR is just $X_{\mathrm{pert}}(I_s) = e Q |I_s|$.

\begin{figure}[tbp]
  \center{\includegraphics[width=1\linewidth]{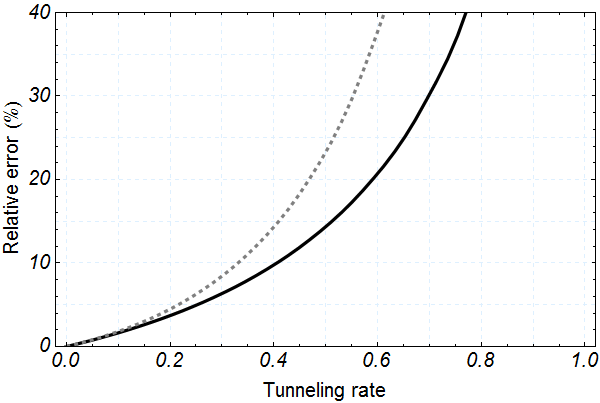}}
  \caption{\textbf{Relative errors of the lowest order perturbative results for the tunneling rate and NtTRR at $\nu = 1/3$ as functions of the exact tunneling rate.} (Color online). The black solid curve is the relative deviation of the perturbative result for the tunneling rate from the exact one. The grey dashed curve is the relative deviation of the perturbative result for the NtTRR from the exact one.}
  \label{fig:TunRateNtTRR_pert_VS_exact_relative_error}
\end{figure}

Fig.~\ref{fig:TunRateNtTRR_pert_VS_exact_relative_error} shows the relative deviation of the lowest order perturbative results for the tunneling rate and NtTRR from the exacts ones. The horizontal axis is the exact tunneling rate. The relation between the exact tunneling rate and $|I_s|/\Xi$ can be seen from Fig.~\ref{fig:Tun_rate_pert_VS_exact}. It is interesting to note that for a given tunneling rate the error of the perturbative NtTRR is generally greater than the error of the perturbative tunneling rate.

While the comparison made in Figs.~\ref{fig:Tun_rate_pert_VS_exact}, \ref{fig:NtTRR_pert_VS_exact}, \ref{fig:TunRateNtTRR_pert_VS_exact_relative_error} gives one an idea of how important the higher order corrections are, the curve representing the exact result in Fig.~\ref{fig:NtTRR_pert_VS_exact} should be taken with a grain of salt in the experimental context. This is because the tunneling amplitude $\Xi$ in a real experiment exhibits a non-universal dependence on $I_s$. It is not untypical that experimentalists work in the regime of constant tunneling rate (see, e.g., \cite{HeiblumExp}). As can be seen from Fig.~\ref{fig:Tun_rate_pert_VS_exact}, this regime corresponds to the ratio $|I_s|/\Xi$ being constant.

\begin{figure}[tbp]
  \center{\includegraphics[width=1\linewidth]{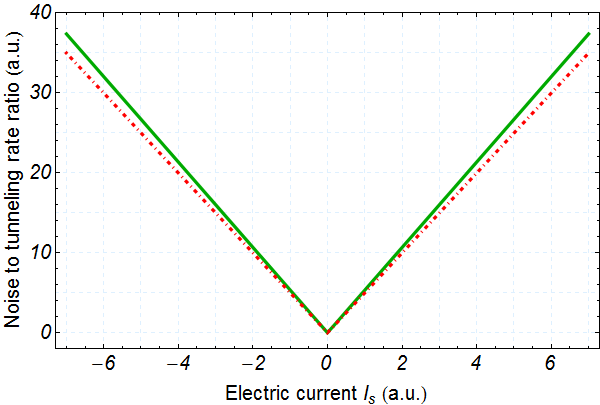}}
  \caption{\textbf{Noise to tunneling rate ratio at $\nu = 1/3$. Perturbative answer vs exact answer in the regime of constant tunneling rate.} (Color online). The red dot-dashed curve is the exact NtTRR plotted using the Eqs.~\eqref{tun_rate_exact_Laughlin_1}--\eqref{noise_exact_Laughlin_2} for $\Xi = 0.5 |I_s|$. The green solid curve is the lowest order perturbation theory answer for the NtTRR, which can be obtained by taking only the first term in the sums in Eqs.~\eqref{tun_rate_exact_Laughlin_1}, \eqref{noise_exact_Laughlin_1}. We remind the reader that the system temperature is equal to $T_0 = 0$.}
  \label{fig:NtTRR_pert_VS_exact_constTR}
\end{figure}

\begin{figure}[tbp]
  \center{\includegraphics[width=1\linewidth]{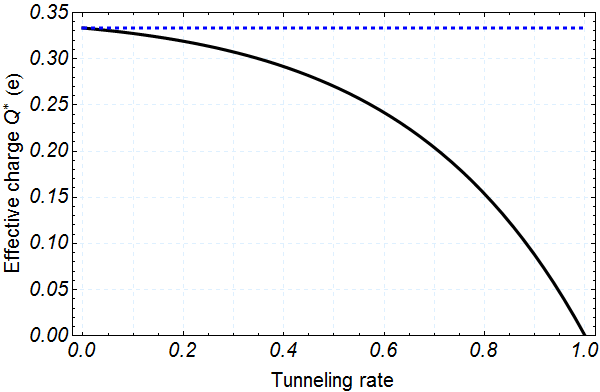}}
  \caption{\textbf{Dependence of the effective charge $Q^*$ on the tunneling rate for $\nu = 1/3$.} (Color online). The black solid curve is the dependence of the effective charge $Q^*$ found from the exact NtTRR on the tunneling rate. The blue dashed line shows the value of the true charge $Q = 1/3$ of the tunneling quasiparticle. We remind the reader that the system temperature is equal to $T_0 = 0$.}
  \label{fig:effective_charge_exact}
\end{figure}

Fig.~\ref{fig:NtTRR_pert_VS_exact_constTR} shows the comparison of the perturbative and the exact answers for the noise to tunneling rate ratio for $|I_s|/\Xi = 2$. Since the temperature $T_0 = 0$, the perturbative answer for the NtTRR is just $X_{\mathrm{pert}}(I_s) = e Q |I_s|$. The exact answer in the regime $|I_s|/\Xi = \mathrm{const}$ is equal to $X_{\mathrm{exact}}(I_s) = e Q^{*} |I_s|$. So the exact answer differs from the perturbative one by the gradient value determined by the "effective charge" $Q^{*}$. As can be seen from Fig.~\ref{fig:effective_charge_exact}, in the limit of infinitely small tunneling rate the effective charge coincides with the true charge of the tunneling quasiparticle: $Q^{*} \rightarrow Q = 1/3$. However, at non-zero tunneling rate the charges do not coincide: $Q^{*} < Q$.

Although, at the moment we are not able to estimate the deviation of the perturbative answer for NtTRR from the exact one at non-zero temperature, the observation that has just been made allows us to formulate some qualitative conditions for the applicability of formula \eqref{Noise_Tun_asymp_full_single_charge}. Namely, one can compare the difference between the answers at zero temperature $e (Q - Q^{*}) |I_s|$ at maximum value of $|I_s|$ that is going to be used with the term $e I_0 (2-8\delta)/\pi = (2-8\delta) k_B T_0 \nu e^2/h$ in Eq.~\eqref{Noise_Tun_asymp_full_single_charge}, where $T_0$ is the system temperature.

For example, at $T_0 = 10\ \mathrm{mK}$ for $|I_s|/\Xi = 2$ (which corresponds to the tunneling rate $r \approx 26\%$) at $I_s = 1\ \mathrm{nA}$ for Lauglin quasiparticle ($Q = 1/3$, $\delta = 1/6$) the term containing $\delta$ is about three times smaller than the error $e (Q - Q^{*}) |I_s|$. Therefore, finding the scaling dimension of the Laughlin quasiparticle with the help of Eq.~\eqref{Noise_Tun_asymp_full_single_charge} is not possible under these experimental conditions.

For quite typical experimental values of $T_0 = 30\ \mathrm{mK}$ and $I_s = 1\ \mathrm{nA}$ the error term does not exceed $(2-8\delta) k_B T_0 \nu e^2/h$ for $r \le 27\%$ and does not exceed $0.1 \times (2-8\delta) k_B T_0 \nu e^2/h$ for $r \le 4\%$. When $e (Q - Q^{*}) |I_s|$ is 10 times smaller than the term containing $\delta$, one can hope to find $\delta$ with a reasonably small error. Thus, if the quality of the experimental data at $r \approx 4\%$ is high enough, it should be possible to find $\delta$ reasonably accurately (with the systematic relative error $\approx$ 10-20\% due to (a) difference between the exact answer and the perturbative one and (b) difference between the perturbative answer and its large-$I_s$ asymptotic behaviour) by fitting the experimental data for NtTRR with Eq.~\eqref{Noise_Tun_asymp_full_single_charge}. One can eliminate the second source of systematic error by using the full formula~\eqref{Noise_Tun_analyt_single_charge} instead of the asymptotic expression ~\eqref{Noise_Tun_asymp_full_single_charge}.

Apart from that, the deviation of the effective charge $Q^{*}$ from the quasiparticle charge $Q$ at higher values of the tunneling rate $r$ gives an opportunity to further check the edge model and the tunneling contact model at $\nu = 1/3$.

\section{\label{Sect_Discussion}DISCUSSION}

\subsection{\label{SubSect_ExpConditions}What experimental conditions are necessary for successful extraction of the scaling dimension?}

Now we discuss the possibility to extract the scaling dimension from real experimental data. The result \eqref{Noise_Tun_asymp_full} shows that it is, in principle, possible to extract the scaling dimension of the tunneling quasiparticles from experimental data on noise to tunneling rate ratio without knowing fully the specific edge theory. However, there are a few practical aspects which should be discussed.

First of all, the parameters $\kappa_i$ related to the quasiparticles' tunneling amplitudes depend on the current $I_s$ in a non-universal way. Therefore, Eq.~\eqref{Noise_Tun_asymp_full} is not useful in the case of several different quasiparticle charges as the gradient of the leading term depends on $I_s$. Therefore, we concentrate on the case when all the charges of the quasiparticles contributing to tunneling are equal: $Q_i = Q$. Examples of such states include the Laughlin series, the Moore-Read Pfaffian, Jain's $\nu = 2/5$ state etc. In this case one can use either Eq.~\eqref{Noise_Tun_asymp_full_single_charge} and extract the scaling dimension from large-$I_s$ asymptotic behavior of the NtTRR, or Eq.~\eqref{Noise_Tun_analyt_single_charge} and analyse the whole dependence of the NtTRR on $I_s$.

Second, we would like to emphasize that while the term containing $\delta$ in Eq.~\eqref{Noise_Tun_asymp_full_single_charge} is proportional to the system temperature $T_0$, it is distinguishable from the Johnson-Nyquist noise $S_{\mathrm{Nyquist}}(0) = \nu T_{0}(2\pi)$. Indeed, the full noise
\eq{S(0) = S_{\mathrm{Nyquist}}(0) + r X(I_s).}
Therefore, the terms we are interested in depend on the tunneling rate $r$, while the Nyquist noise does not. Moreover, expression \eqref{Noise_Tun_asymp_full_single_charge} is the large-$I_s$ asymptote of the NtTRR, while the Nyquist noise is present independently of $I_s$, and at $I_s = 0$ is the only noise present:
\eq{X(I_s = 0) = 0 \Rightarrow S(0)\left|_{I_s = 0}\right. = S_{\mathrm{Nyquist}}(0).}
Thus, the Nyquist noise can be subtracted at $I_s = 0$ without confusing it with the second term in Eq.~\eqref{Noise_Tun_asymp_full_single_charge}.

There is, however, a technical problem. The term we are interested in a constant offset of the large-current asymptote of $S(0)$ which is much smaller than the Nyquist noise one has to subtract. Therefore, a small error in the subtracted value of $S_{\mathrm{Nyquist}}(0)$ can lead to a huge error in the determined $\delta$. It this context, using the full expression for the NtTRR \eqref{Noise_Tun_analyt_single_charge} rather than its asymptote \eqref{Noise_Tun_asymp_full_single_charge} can be advantageous.

The third issue is that the dynamics of the system changes near a characteristic energy scale in the FQHE system. Namely, there is a bulk gap $\Delta$. As the typical energies of the system exceed $\Delta$, bulk dynamics starts being involved. Thus, one should restrict oneself to
\eq{
|I_s| \lesssim \nu \frac{e}{h} \pi \Delta.
}
Deviations from our theory can be expected beyond this threshold.\footnote{In some recent works, e.g. \cite{Ferraro2008,Ferraro,Carrega2011,Carrega2012}, it is argued that the neutral mode energy cutoffs, which can be smaller than the bulk gap $\Delta$, should be introduced to explain some of the observed effects. In that case these cutoffs should also be taken into account. We discuss this issue in more detail in subsection~\ref{SubSect_Relations}.}

\begin{figure}[tbp]
  \center{\includegraphics[width=1\linewidth]{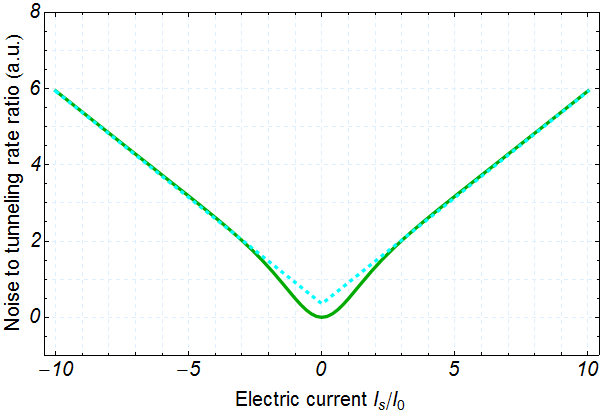}}
  \caption{\textbf{Noise to tunneling rate ratio vs its asymptotic behaviour}. (Color online). The green solid curve is the NtTRR \eqref{Noise_Tun_analyt_single_charge} for $Q = 1/3$, $\delta = 1/6$. The cyan dashed curve is the large-$I_s$ asymptote of the NtTRR \eqref{Noise_Tun_asymp_full_single_charge} for the same values of $Q$ and $\delta$. The asymptote almost coincides with the original curve for $|I_s| \ge 3 I_0$.}
  \label{fig:NtTRR_full_VS_asymp}
\end{figure}

The fourth issue is the lower validity bound for the asymptotic expression \eqref{Noise_Tun_asymp_full_single_charge}. Fig.~\ref{fig:NtTRR_full_VS_asymp} shows the comparison of the NtTRR \eqref{Noise_Tun_analyt_single_charge} against its asymptotic behaviour \eqref{Noise_Tun_asymp_full_single_charge} for $Q = 1/3$ and $\delta = 1/6$. These parameters correspond to the most relevant quasiparticle of the simplest $\nu = 1/3$ edge model. As one can see, for $|I_s| \ge 3 I_0$ the exact NtTRR and its large-$I_s$ asymptote almost coincide.

To estimate how close the asymptote and the original curve are we have done some fitting. Namely, we took part of a part of the original curve with $|I_s|$ between $\alpha I_0$ and $10 I_0$ and fitted it with \eqref{Noise_Tun_asymp_full_single_charge} using $Q$ and $\delta$ as fitting parameters. For $\alpha \ge 3$ the fitted charge and scaling dimension deviate from their correct values by less than 1\% and 11\% respectively. This gives an idea of how accurate the estimates of $Q$ and $\delta$ obtained from fitting experimental data with formula \eqref{Noise_Tun_asymp_full_single_charge} can be if there are no other sources of errors.

Thus, one can use the asymptotic expression \eqref{Noise_Tun_asymp_full_single_charge} for $|I_s| \gtrsim \alpha I_0$, where $\alpha$ is on the order of 1. The exact value of the multiplier $\alpha$ depends on the values of $Q$ and $\delta$. Of course, this issue does not arise if one uses the full expression~\eqref{Noise_Tun_analyt_single_charge}.

Note that the greater is $I_0$ the more significant is the term containing the scaling dimension in Eq.~\eqref{Noise_Tun_asymp_full_single_charge}. At the same time, the less is the interval $\nu \frac{e}{h} \pi \Delta \gtrsim |I_s| \gtrsim \alpha I_0$. Thus, the choice of the system temperature should be a matter of trade-off between these two restrictions in order to allow as good determining of the scaling dimension $\delta$ as possible.

Fifth. The expressions \eqref{Noise_Tun_asymp_full_single_charge}, \eqref{Noise_Tun_analyt_single_charge} are valid only when the contribution of less relevant quasiparticles (with greater scaling dimensions) to the tunneling processes can be neglected. Otherwise the corrections due to less relevant quasiparticles can hinder finding the scaling dimension using the large-$I_s$ NtTRR behaviour. Unfortunately, there are no known reliable ways to estimate theoretically how significant these corrections are. However, general theoretical arguments, as well as recent Monte Carlo simulations \cite{HuntingtonCheianov_TunnAmplitudeMonteCarlo}, show that the tunnelling amplitude of a quasiparticle with scaling dimension $\delta$ is proportional to $(L/l_{B})^{-2\delta}$, where $l_{B}$ is the magnetic length, and $L$ is the edge length. The typical experimental values of these parameters are $l_B \approx 10\ \mathrm{nm}$, $L \approx 10\ \mathrm{\mu m}$, suggesting that the contribution of less relevant quasiparticles can usually be neglected. However, to be on the safe side, one can estimate them in practice by comparing experimental data with different possible theoretical answers for NtTRR (the answers including and not including less relevant quasiparticles).

The sixth issue is related to measurement errors. Scaling dimension enters Eq.~\eqref{Noise_Tun_asymp_full_single_charge} as a subleading term. Thus, finding the scaling dimension demands a very high quality experimental data with very small statistical errors. The NtTRR errors can be made less significant by using greater values of the tunneling rate. This, however, worsens the accuracy of theoretical result \eqref{Noise_Tun_asymp_full_single_charge} which was derived perturbatively in the limit of small tunneling rate. Therefore, the choice of the strength of tunneling in experimental data should be balanced between worsening the applicability of the theory and improving the quality of data for NtTRR.

The latter observation brings up the seventh issue. The theoretical result \eqref{Noise_Tun_asymp_full_single_charge} was derived perturbatively in the limit of weak tunneling of the quasiparticles. One can reasonably expect that if the tunneling rate is about, e.g., 10\% the next perturbative correction to (and the inaccuracy of) the NtTRR should also be about 10\%. While such an inaccuracy would bring about an error of the same order to the determined charge $Q$, the effect on the subleading term may be much more significant. This imposes a strong restriction on the value of the tunneling rate as is elaborated in section \ref{Sect_ExactResults}. There we find that for $\nu = 1/3$ the error of the perturbative formulae is smaller than one could expect a priori. In particular, for $\nu = 1/3$ and typical experimental parameters one needs the tunneling rate $r \lesssim 5\%$ in order to introduce no more than $10\%$ error to the $\delta$-containing term.

Finally, we have to comment on the issue of finite frequency in the real noise measurements. Indeed, in typical experiments (see, e.g., Refs.~\cite{HeiblumExp, Heiblum_2013_LaughlinFilling_NeutrMode, Chung2003, FracChargeObs_Heiblum, FracChargeObs_Glattli, Dolev2010}) the current noise is measured at finite frequencies $\omega \lesssim 1 \mathrm{MHz}$ in order to reduce the influence of $1/f$ noise. The finite frequency of noise introduces corrections to our formulas. However, we estimate those corrections to be 2 to 3 orders of magnitude smaller than the $\delta$-containing term at typical experimental conditions. Therefore, we neglect those corrections.

To summarize, the NtTRR \eqref{Noise_Tun_analyt_single_charge} and its large-$I_s$ asymptotic behaviour \eqref{Noise_Tun_asymp_full_single_charge} can be used to find the scaling dimension of the most relevant quasiparticle. One should, however, take care to choose the appropriate parametric regime in order to reduce errors. With ideal experimental data at realistic experimental conditions it should be possible to extract the scaling dimension to about 10-20\% accuracy.

\subsection{\label{SubSect_Relations}Relation to other developments in the field}

We performed our analysis with some assumptions regarding the physics of the FQHE edge theory and the QPC. However, both the experimental data and theoretical studies suggest that a more complicated picture than we used is necessary to analyze the tunneling experiments. Here we comment on how such complications are going to affect our proposed method for finding the most relevant quasiparticle scaling dimension.

First of all, we assumed that the non-universal physics at the QPC involves the tunneling amplitudes only. However, several works proposed mechanisms that would lead to a renormalization of the tunneling operator scaling dimension in a sample-dependent way (and, possibly, also dependent on the tunable system parameters such as bias voltage) \cite{Rosenow2002, Papa2004, Carrega2011}. If the renormalization of the scaling dimension does happen in the system in a bias voltage independent way, then our method is still applicable but the determined scaling dimension will be the renormalized one. This would open up a possibility to study the non-universal physics leading to the renormalization in a quantitative way. If, conversely, the renormalization strength changes with the bias voltage (or with the current $I_s$, which is essentially the same), then our method is no longer applicable.

Second, there have been experiments reporting that the effective charge of tunneling quasiparticles (which is the leading term in the NtTRR large-$I_s$ asymptote) is not constant but is rather a function of the system's temperature, bias voltage or other parameters \cite{Chung2003, HeiblumExp, Dolev2010, HeiblumForFerraro}. Several works proposed theoretical explanations for this.

One explanation involves introducing energy cutoffs to the transport channels of the FQHE edge, such that the cutoff of the neutral mode(s) is smaller than the cutoff of the charged mode \cite{Ferraro2008,Ferraro}. Then as the energy scale (temperature or bias voltage) becomes greater than the neutral mode cutoff, the edge dynamics changes and so do the quasiparticles' scaling dimensions, leading to a different quasiparticle contributing most to tunneling. In the context of our work this implies that if the system is in either of the regimes, one can use our method to extract the scaling dimension of the most relevant quasiparticle in the appropriate regime. If the system energy scale is around the neutral mode(s) cutoff, then the system is in the transition between the two regimes, and our calculation should be modified. The typical values of the cutoffs $\omega_\mathrm{n}$ proposed for different filling factors vary from $50 \mathrm{mK}$ to $200 \mathrm{mK}$ \cite{Ferraro, Ferraro2008, Carrega2011}. For the typical experimental parameters of temperature and bias voltage it should be possible to investigate the low-energy regime, where the cutoffs have little influence, for $\omega_\mathrm{n} \gtrsim 200 \mathrm{mK}$ and the high-energy regime, where the neutral modes are saturated, for $\omega_\mathrm{n} \lesssim 50 \mathrm{mK}$. The values in between, however, would mean that one has to deal with the intermediate regime. We note that whether the use of this double-cutoff model is appropriate has to be decided on a case by case basis.

Another type of explanation involves considering a reconstructed (with additional transport channels) edge which behaves in a different way at different energies due to edge disorder \cite{MeirGefen23EdgeReconstruction}. However, between the transition energies the edge can be described by a model of the type we have considered in this paper. Therefore, with our proposed method one can extract the scaling dimension of the most relevant quasiparticle in the corresponding regime of the edge theory.

Summarizing, there has been a number of theoretical works which consider more complicated theories of what happens in the tunneling experiments in order to explain some experimental observations. The model we have considered in this paper is simpler and does not incorporate all the elements of the proposed theories. However, in many cases our model can describe some of the regimes of the proposed theories, and thus our method can be used to extract the scaling dimension of the most relevant quasiparticle in the appropriate regime.

\section*{CONCLUSIONS}

We propose a method for finding the scaling dimension of the most relevant quasiparticle at a quantum Hall edge using tunneling current and tunneling current noise measurements. The advantages of the method are (a) reduced sensitivity to the non-universal physics of tunneling contacts (compared to methods based solely on tunneling current measurements), (b) a certain degree of model independence. By comparing our perturbative results with the exact results of Ref.~\cite{Saleur2} in the case of $\nu = 1/3$ we find that our method should be applied for small enough tunneling rates $r \lesssim 5\%$.

Using the exact solution of Ref.~\cite{Saleur2} at $\nu = 1/3$ for higher tunneling rates, we find that the effective charge $Q^{*}$ which can be found from an experiment using standard perturbative formulae deviates from the true charge of the most relevant quasiparticle $Q$. We propose to measure and study this difference in order to check the minimal $\nu = 1/3$ edge model and the tunneling contact model.

\section*{Acknowledgements}

We would like to thank Oles' Shtanko, Anna Grivnin and Moty Heiblum for useful discussions.

The research leading to these results has received funding from the European Research Council under the European Union's Seventh Framework Programme (FP7/2007-2013) / ERC grant agreement No 279738 - NEDFOQ.

\appendix

\section{\label{App_HowGeneralNoiseAnswers}How general are the answers of section~\ref{Sect_ScalDimFromAsymp}?}

In section~\ref{Sect_ScalDimFromAsymp} we obtained formulae \eqref{tunn_coeff_expr}-\eqref{j_s} for the tunneling rate and the tunneling current noise within the leading order perturbation theory in tunneling Hamiltonian. They were obtained for Abelian models. However, these formulae and the calculations leading to them are straightforward to generalize to a much wider class of edge theories including some of the non-Abelian ones.


In section~\ref{Sect_ScalDimFromAsymp} we considered a general Abelian quantum Hall (QH) with the only restriction that all the modes that carry electric charge have the same chirality $\chi_{i}$. One typically expects this to be the case in the quantum Hall effect. If a theory contains counter-flowing charged modes, in the low-energy limit it can become a theory with a set of charged modes propagating in one direction and a set of neutral modes (possibly, with different directions of propagation) according to the mechanism described in Refs.~\cite{KaneFisher, KaneFisherPolchinski}.

In the case of such theories one can show that the formulae \eqref{tunn_coeff_expr}-\eqref{j_s} hold for tunneling of the quasiparticles with $\delta < 1/2$. For tunneling of the quasiparticles with $\delta \geq 1/2$ only the formulae \eqref{G_i}, \eqref{F_TT}, \eqref{F_0T} should be modified with the terms cancelling divergencies of the integrals at $t \rightarrow 0$ similar to the $\varepsilon^{1-4\delta}$ term in Eq.~\eqref{F_TT}.

A more general class of QH edge theories is where the charged sector is still described in terms of free bosons like in Abelian theories, while the neutral sector is described in terms of a more complicated model~--- some conformal field theory (CFT). Perhaps, the most famous example of such a model corresponds to the Moore-Read Pfaffian state. A general scheme for construction of such models is described in Ref.~\cite{Frohlich_CosetHall}. For more details on CFT see Ref.~\cite{Francesco_CFT}. For the purposes of the present work it suffices to say that the leading order perturbation theory results \eqref{tunn_coeff_expr}-\eqref{j_s} hold for this class of models as well as they do for the Abelian ones.

We emphasize that the phenomenological assumptions regarding the interaction of the Ohmic contacts with the edge are important for the derivation of formulae \eqref{tunn_coeff_expr}-\eqref{j_s}. Most importantly, we assume that (a) all the excitations of charged and neutral modes are fully absorbed by the Ohmic contacts they flow into and (b) that the lower edge temperature does not depend on the current $I_s$. The latter one is rarely mentioned but is crucial for the results of the present work.

So, the formulae \eqref{tunn_coeff_expr}-\eqref{j_s} are valid for a wide class of typical Abelian and non-Abelian FQHE edge models.

\section{\label{App_LargeCurrAsympDeriv}Large-$I_s$ asymptotic behaviour of the noise to tunneling rate ratio}

Consider the large-$I_s$ limit of Eq.~\eqref{G_i}. For $|j_s| \gg 1$ one gets
\nmq{
\label{G_i_asymp}
G_i = \frac{j_s}{|j_s|} Q_i^{4\delta} |j_s|^{4\delta - 1} \sin{2\pi\delta} \times\\
 \left(\int\limits_{0}^{\infty}dx \frac {\sin{x}}{x^{4\delta}} + O\left(\frac{1}{Q_i^2 j_s^2}\right) \right) =\\
  = \frac{j_s}{|j_s|} \frac{\pi }{2 \Gamma{(4\delta)}} Q_i^{4\delta} |j_s|^{4\delta - 1} \left(1 + O\left(\frac{1}{Q_i^2 j_s^2}\right) \right),
}
where $\Gamma{(x)}$ is the Euler gamma function.

Similarly, for Eqs.~\eqref{F_TT}, \eqref{F_0T}, \eqref{F_i} in the limit $|j_s| \gg 1$ one gets
\nmq{
\label{F_TT_asymp}
F^{TT}_{i} = \frac{\pi}{2 \Gamma{(4\delta) \cos{2 \pi \delta}}} Q_i^{4\delta + 1} |j_s|^{4\delta - 1} \times\\
 \left(1 + O\left(\frac{1}{Q_i^2 j_s^2}\right) \right),
}
\nmq{
\label{F_0T_asymp}
F^{0T}_{i} = \frac{\pi (4\delta - 1)}{2 \Gamma{(4\delta) \sin{2 \pi \delta}}} Q_i^{4\delta} |j_s|^{4\delta - 2} \times\\
 \left(1 + O\left(\frac{1}{Q_i^2 j_s^2}\right) \right),
}
\nmq{
\label{F_i_asymp}
F_i = \frac{\pi}{2 \Gamma{(4\delta)}} Q_i^{4\delta} |j_s|^{4\delta - 2} \times\\
 \left(Q_i |j_s| + \frac{2-8\delta}{\pi} + O\left(\frac{1}{Q_i^2 j_s^2}\right) \right).
}

Using Eqs.~\eqref{tunn_coeff_expr}, \eqref{noise_expr}, \eqref{noise_tun}, \eqref{G_i_asymp}, \eqref{F_i_asymp} one finally gets the asymptotic expression for the NtTRR \eqref{Noise_Tun_asymp_full}:
\nmq{
\label{Noise_Tun_asymp_full_Appendix}
X(I_s)\left|_{|j_s| \gg 1}\right. = \frac{\tilde{S}(0)}{r} = e I_s \frac{\sum_{i} \theta_i F_i}{\sum_{i} \theta_i G_i} =\\
 e I_s \frac{\sum_i \theta_i \left(Q_i^{4\delta+1} |j_s| + Q_i^{4\delta} \frac{2-8\delta}{\pi} + O\left(\frac{1}{Q_i |j_s|}\right) \right)}{j_s \sum_i \theta_i Q_i^{4\delta} \left(1 + O\left(\frac{1}{Q_i^2 j_s^2}\right) \right)} =\\
  e |I_s| \frac{\sum_i \theta_i Q_i^{4\delta+1}}{\sum_i \theta_i Q_i^{4\delta}} + e I_0 \frac{2-8\delta}{\pi} + O\left(|j_s|^{-1}\right).
}

\section{\label{App_AnalyticAnswerForNtTRR_Deriv}Analytic expressions for the noise to tunneling rate ratio}

For the following derivation we need several facts about Euler beta function $\mathrm{B}{(x,y)}$ and Euler gamma function~$\Gamma{(x)}$.
\begin{gather}
\label{GamGamSin}
\Gamma{(x)}\Gamma{(1-x)} = \frac{\pi}{\sin{\pi x}},\\
\label{GamComplConj}
\Gamma{(\bar{x})} = \overline{\Gamma{(x)}},\\
\label{BetaViaGamma}
\mathrm{B}{(x,y)} = \frac{\Gamma{(x)}\Gamma{(y)}}{\Gamma{(x+y)}},\\
\label{BetaIntegralSinh}
2^{4\delta - 1} \mathrm{B}{\left(1-4\delta, \frac{\alpha}{2} + 2 \delta\right)} = \int_{0}^{\infty} dt \frac{e^{-\alpha t}}{(\sinh{t})^{4 \delta}}.
\end{gather}
The bars in the second equation denote complex conjugation. The last identity holds for $\delta < 1/4$ and $\mathrm{Re}\left[\alpha\right]/2 + 2 \delta > 0$, where $\mathrm{Re}{\left[...\right]}$ denotes taking of the real part. However, it can be analytically continued beyond these restrictions.

Using Eqs.~\eqref{GamGamSin}-\eqref{BetaIntegralSinh}, one can get the following analytic expressions for the functions defined in Eqs.~\eqref{G_i}, \eqref{F_TT}:
\eq{
\label{G_i_analyt}
G_i = \frac{Q_i 2^{4\delta - 2}}{\Gamma{(4\delta)}} \left|\Gamma{\left(2\delta + \frac{i Q_i j_s}{2}\right)}\right|^2 \sinh{\frac{\pi Q_i j_s}{2}},
}
\nmq{
\label{F_TT_analyt}
F^{TT}_{i} =\\
 \frac{Q_i^2 2^{4\delta-2}}{\Gamma{(4\delta)} \cos{2 \pi\delta}} \left|\Gamma{\left(2\delta + \frac{i Q_i j_s}{2}\right)}\right|^2 \cosh{\frac{\pi Q_i j_s}{2}}.
}

For the function defined in Eq.~\eqref{F_0T}, noting that\footnote{An interesting relation between this fact and the Ward identity arising due to the conservation of electric charge was noted in Ref.~\cite{Simon_QH_NonEqNoise_FDT}.}
\eq{F^{0T} = \frac{1}{\sin{2 \pi \delta}} \frac{\partial}{\partial j_s} G_i,}
one gets
\nmq{
\label{F_0T_analyt}
F^{0T}_{i} =\\
 \frac{Q_i^2 2^{4\delta - 2}}{\Gamma{(4\delta)} \sin{2 \pi\delta}} \left|\Gamma{\left(2\delta + \frac{i Q_i j_s}{2}\right)}\right|^2 \sinh{\frac{\pi Q_i j_s}{2}} \times\\
 \times \left(\frac{\pi}{2}\coth{\frac{\pi Q_i j_s}{2}} - \mathrm{Im}\left[\psi{\left(2\delta + \frac{i Q_i j_s}{2}\right)}\right] \right),
}
where the digamma function $\psi{(x)} = (\ln{\Gamma{(x)}})'$ is the logarithmic derivative of the Euler gamma function $\Gamma{(x)}$, and $\mathrm{Im}{\left[...\right]}$ denotes taking of the imaginary part.

Thus, for $F_i$ defined in Eq.~\eqref{F_i} we have
\nmq{
\label{F_i_analyt}
F_i = \frac{Q_i^2 2^{4\delta - 1}}{\pi \Gamma{(4\delta)}} \left|\Gamma{\left(2\delta + \frac{i Q_i j_s}{2}\right)}\right|^2 \times\\
 \sinh{\frac{\pi Q_i j_s}{2}} \mathrm{Im}{\left[\psi{\left(2\delta + \frac{i Q j_s}{2}\right)}\right]}.
}

Using Eqs.~\eqref{tunn_coeff_expr}, \eqref{noise_expr}, \eqref{noise_tun}, one straightforwardly gets the analytic expression for the noise to tunneling rate ratio $X(I_s)$. In the case of coinciding charges of all the quasiparticles participating in tunneling this expression simplifies significantly leading to the result \eqref{Noise_Tun_analyt_single_charge}.

\bibliography{Thesis}
\end{document}